\documentclass[12pt]{article}
\pdfoutput=1
\usepackage{lutypaper,bm,mathdots}
\numberwithin{equation}{section} 

\begin{document}

\newcommand{\tc}{{\ensuremath{\tau_\text{c}}}}
\newcommand{\rc}{{\ensuremath{r_{\text{c}}}}}
\renewcommand{\vec}[1]{\mathbf{#1}}
\newcommand{\xhat}{\hat{\vec{x}}}
\newcommand{\yhat}{\hat{\vec{y}}}
\newcommand{\zhat}{\hat{\vec{z}}}
\newcommand{\rhat}{\hat{\vec{r}}}

\begin{titlepage}
\title{Radiation Reaction, \\
Over-reaction, and Under-reaction}

\author{Mario D'Andrea,\ \ Markus A. Luty,\ \ Christopher B. Verhaaren}
\address{Center for Quantum Mathematics and Physics (QMAP)\\
University of California, Davis}

\begin{abstract}
The subject of radiation reaction in classical electromagnetism
remains controversial over 120 years after the pioneering work of Lorentz.
We give a simple but rigorous treatment of the subject at the textbook level that 
explains the apparent paradoxes that are much discussed in the literature on
the subject.
We first derive the equation of motion of a charged particle from 
conservation of energy and momentum, which includes the self-force term.
We then show that this theory is unstable if charged particles are pointlike:
the energy is unbounded from below, and charged particles self-accelerate
(`over-react') due to their negative `bare' mass.
This theory clearly does not describe our world,
but we show that these instabilities are absent if the particle has a finite size 
larger than its classical radius.
For such finite-size charged particles, the effects of radiation reaction can be computed
in a systematic expansion in the size of the particle.
The leading term in this expansion 
is the reduced-order Abraham-Lorentz equation of motion, which 
has no stability problems.
We also discuss the apparent paradox that a particle with constant acceleration radiates,
but does not suffer radiation reaction (`under-reaction').
Along the way, we introduce the ideas of renormalization and effective theories,
which are important in many areas of modern theoretical physics.
We hope that this will be a useful addition to the
literature that will remove some
of the air of mystery and paradox surrounding the subject.
\end{abstract}

\end{titlepage}

\section{Introduction and Overview\label{s.intro}}
A charged particle that emits electromagnetic radiation must lose energy and momentum.
This `radiation reaction' is a textbook subject that is often 
included in advanced undergraduate and graduate courses in electrodynamics.
Radiation reaction is a tiny effect on the instantaneous motion of a charged particle.
As we discuss in detail below, the correction to the acceleration of an electron
due to radiation reaction is of order
\[
\frac{\de a}{a} \sim \frac{r_e}{c T},
\]
where $r_e \sim 3\times10^{-13}$~cm is the classical radius of the electron,
$c$ is the speed of light, and $T \sim a/\dot{a}$ is the time scale for the
change in acceleration.
The classical radius of the electron is smaller than the length scale at which
quantum effects become important, and so the instantaneous effis a rapidects of radiation 
reaction are always small whenever classical electrodynamics is valid.
It is true that the 
effects of radiation reaction can build up over time and become important 
(for example in antennas), but in these situations the effects of radiation reaction
can be taken into account using simple energy conservation arguments.
Although there is no disagreement about any experimentally measurable effects
of radiation reaction reaction, the subtleties in the theory have led to an
immense and still-growing literature on the subject.%
\footnote{For a survey of the literature, see \cite{McDonald}.}

There is near-universal agreement that radiation reaction is 
described by the Abraham-Lorentz (AL) force, 
which is proportional to the time derivative of the acceleration.
However, this equation of motion appears to predict that electrons should 
self-accelerate to nearly the speed of light on a time scale $r_\text{e}/c$.
This `over-reaction' is in in gross disagreement with observation.
Another paradoxical consequence of the AL force is that
a charged particle undergoing constant acceleration has no radiation reaction
despite the fact that it radiates; 
it apparently `under-reacts.'
Discussions of these apparent paradoxes account for much of the literature
on radiation reaction.

In this article, we take a fresh look at this subject.
As might be expected in a problem that is  
over 120 years old and that has attracted
the attention of some of the century's most prominent physicists
\cite{Poincare1891,Lorentz1892,Poincare1894,Planck1897,Abraham1905,Sommerfeld1905,Pauli1921,Dirac:1938nz,Wheeler:1949hn,Coleman:1961zz}, 
most of the conclusions of this paper can be found somewhere in the literature.
The main novelty of our approach is that we consistently expand all effects
in the size of the particle using the ideas of renormalization and effective
theory.
These are powerful tools of modern theoretical physics,
and our treatment introduces them in an elementary context.
Our aim is to give a complete and logically coherent account of radiation
reaction at the textbook level using these ideas.
We hope that this will be a useful addition to the literature, 
though we dare not hope that it definitively 
puts to rest this `perpetual problem' \cite{Ginsburg}.

The main points of our treatment are as follows:
\begin{enumerate}
\item 
Energy and momentum conservation determine the electromagnetic forces
on a classical charged point particle, including the effects of radiation reaction.
This yields the Lorentz force with the addition of the AL term.
\item 
Short-distance divergences associated with a point charge are
regulated by replacing it with a charge of finite size $r_0$.
This model must be chosen to satisfy local conservation of energy as
well as special relativity.%
\footnote{For the cogniscenti, this requires additional 
`Poincar\'e stress'~\cite{Poincare1905} terms in the energy-momentum tensor, 
which resolve the so-called `$4/3$ problem.'}
Physical quantities are finite and unambiguous in the $r_0 \to 0$ limit,
provided that the mass of the particle is taken to depend on $r_0$.
This `renormalization' procedure is conceptually similar to the 
procedure used in quantum field theory to eliminate short-distance divergences.
\item 
In the $r_0 \to 0$ limit, the theory suffers from catastrophic instabilities.
The energy is unbounded from below because charged particles have negative
mass, and the generic behavior of a charged particle
is a rapid self-acceleration to nearly the speed of light
while emitting electromagnetic radiation.
\item These instabilities can be avoided only if 
the particle has a physical size $r_0 > \rc$, 
where $\rc = q^2/4\pi m$ is the classical radius associated with the charge of the particle. 
If $r_0 \gg \rc$, but $r_0$ is  smaller than the other relevant physical length scales in the 
problem, one can define an `effective theory' 
that gives a systematic expansion of physical quantities in a power series in $r_0$.
In this expansion the leading effects of radiation reaction are 
described by the reduced-order AL equation.
\item
This `effective AL equation' correctly predicts that radiation reaction is always
a small effect; in particular, the runaway solutions are absent.
The equation predicts that the energy lost to radiation is equal to the change in
the kinetic energy of the particle for external forces that are periodic, or 
that turn off at early and late times.
In particular, this holds for particles that experience constant acceleration for a
finite time, which we use as an example to illustrate the results.
\end{enumerate}
The final results are very simple, and it is our hope that our work removes some
of the mystery surrounding the subject of radiation reaction, inspiring
confidence in the correctness of the effective AL equation to describe radiation reaction.

\section{Forces from Conservation Laws\label{s.derevation}}
We begin by demonstrating that the electromagnetic
forces on a pointlike charged particle can be unambiguously
determined from energy and momentum conservation.
Our assumptions are:
\begin{enumerate}
\item Special relativity is valid.
\item The vacuum Maxwell equations describe the dynamics of electromagnetic
fields away from the charges.
\item
The electromagnetic field of a given charge distribution is given by the standard
retarded solutions.
\item Energy and momentum are conserved, and the energy and momentum of the
electromagnetic fields is given by the standard formulas.
\end{enumerate}
The idea is that electromagnetic interactions exchange
energy and momentum between the particle and the fields.
The conservation of total energy can then be used to 
derive the Lorentz force, as well as the `self force' that accounts
for radiation reaction.
This derivation makes it clear that the apparent paradoxes associated with the 
self-force are not due to violations of energy and momentum conservation.
This idea is not new \cite{Dirac:1938nz}, but it does not seem to have found
its way into standard textbooks on electrodynamics.
We provide an elementary but rigorous derivation that should be accessible
to the advanced undergraduate or beginning graduate student.

To carry out this derivation, we must deal with the fact that the 
electric field of a point charge diverges near the charge, and therefore
the energy and momentum density in the electromagnetic field diverge at
the position of the charge.
We regulate these divergences by introducing a simple model for a finite size charge.
This model is not intended to be realistic, but we expect the details of the model 
to be unimportant in the limit where the size of the particle is smaller than other 
length scales in the problem.
In our model, we assume that the electromagnetic fields turn off at distances
$r < r_0$ in the particle's rest frame.
(Special relativity implies that the shell will not be a sphere in
a reference frame where the charge is not at rest.)
We can think of $r_0$ as the size of the particle: we attribute
all energy and momentum for $r < r_0$ to the `particle,'
while everything outside the shell is governed by classical electromagnetism.
In the limit where $r_0$ is smaller than all other scales
in the problem, this simple model gives well-defined equations of motion
that are independent of $r_0$, as well as the model we chose.
For example, they agree with the predictions of physical models of a
classical extended charge in the limit where the size of the charge is small
\cite{Sommerfeld1905}.

We assume that the particle has a trajectory $\vec{X}(t)$ in the presence of
external electromagnetic fields $\vec{E}_\text{ext}(\vec{r}, t)$, $\vec{B}_\text{ext}(\vec{r}, t)$.
Our goal is to derive an equation for $\vec{X}(t)$ from conservation of
energy and momentum.
The energy and momentum of the particle (everything inside the shell)
is assumed to have the form dictated by special relativity
\[
\eql{Eppart}
\scr{E}_\text{part} = \ga m_0,
\qquad
\vec{p}_\text{part} = \ga m_0 \vec{v},
\]
where $m_0$ is the `bare' mass of the particle,
$\vec{v} = \dot{\vec{X}}$ is the velocity of the particle,
and $\ga = (1 - v^2)^{-1/2}$.
We use units where $c = 1$.
In Section~\ref{s.em}, we will show that the energy
and momentum of the particle
must have additional contributions in order to be consistent with 
special relativity, but these additional contributions do not affect
the argument below.
The change of the energy and momentum of the particle is given by the
flux of energy and momentum through the shell
that surrounds the particle:
\[
\eql{depdt}
\dot{\scr{E}}_\text{part} = -\int \! d\vec{A} \cdot \vec{S},
\qquad
\dot{\vec{p}}_\text{part} = -\int \! d\vec{A} \cdot \vec{T},
\]
where $\vec{S} = \vec{E} \times \vec{B}$ is the Poynting flux, and
\[
\eql{Tdef}
T^{ij} = -E^i E^j - B^i B^j
+ \sfrac12 \delta^{ij} \left(E^2+B^2 \right),
\]
are the spatial components of the electromagnetic energy-momentum tensor.
We use Heavy\-side-Lorentz units.
The minus signs in \Eq{depdt} indicate the integrals give the flux
of energy and momentum \emph{out} of the shell
($d\vec{A}$ points outward).

The total electromagnetic fields are a sum of the external fields and the
fields of the particle.
Outside the shell of radius $r_0$, the fields of the particle are assumed
to be given by the standard retarded 
solutions for a point particle with charge $q$
\[
\vec{E}_\text{part}(\vec{r},t)=&\frac{q}{4\pi}\left( \frac{R}{\vec{R}\cdot\bm{\rho}}\right)^3\left[\left(1-v^2 \right)\bm{\rho}+\vec{R}\times\left( \bm{\rho}\times\vec{a}\right) \right] \biggl|_{t \, = \, t_r},
\label{e.Epoint} \\
\vec{B}_\text{part}(\vec{r},t)=& \frac{\vec{R}(t_r)}{R(t_r)} \times\vec{E}_\text{part}(\vec{r},t).\label{e.Bpoint} 
\]
where $q$ is the charge of the particle,
$\vec{R}(t) \equiv \vec{r} - \vec{X}(t)$ is a vector that points from the
particle to the observation point $\vec{r}$, and we use the abbreviation
$\bm{\rho}(t)\equiv \vec{R}(t)-R(t)\vec{v}(t)$.
The retarded time $t_r$ is determined by
\[
t_r= t-R(t_r).
\label{e.rTime}
\]
We want to compute the change of the particle's energy and momentum at $t = 0$,
given by \Eq{depdt}.
Without loss of generality, we work in the instantaneous rest frame of the particle, 
so that $\vec{v}(0) = \dot{\vec{X}}(0) = 0$.
We also choose the coordinates so that $\vec{X}(0) = 0$.
Because we are in the rest frame, the shell surrounding the particle is 
a sphere of radius $r_0$.
Because $r_0$ is assumed to be smaller than all other scales in
the problem, we only need to know the fields close to the particle.
We can obtain a systematic expansion for the fields in powers of $r$ as follows.
For the external fields, this is simply a Taylor expansion:
\[
\eql{extfieldexpand}
\vec{E}_\text{ext}(\vec{r},0) = \vec{E}_\text{ext}(0,0) + O(r),
\qquad
\vec{B}_\text{ext}(\vec{r},0) = \vec{B}_\text{ext}(0,0) + O(r).
\]
For the fields due to the point charge, we expand 
the motion of the particle for small $t_r$ as
\[
\vec{X}(t_r) = \frac 12 \vec{a}(0) t_r^2 + \frac 16 \dot{\vec{a}}(0) t_r^3 + O(t_r^4).
\]
We can use this expression in Eq.~\eqref{e.rTime} 
to obtain (at $t = 0$)
\[
t_r=-r+\sfrac12t_r^2 \, \vec{a}(0)\cdot\rhat
+\sfrac16t_r^3 \, \dot{\vec{a}}(0)\cdot\rhat+ O(t_r^4)\,.
\]
This can be rewritten as a power series in small $r$ by expanding about the 
leading solution $t_r = r$:
\[
t_r=-r+\sfrac12\vec{a}\cdot\rhat\,r^2-\sfrac16\left[\dot{\vec{a}}\cdot\rhat+\sfrac34a^2+\sfrac94(\vec{a}\cdot\rhat)^2 \right]r^3+ O(r^4).
\]
Using this relation the fields due to the particle at $t = 0$
can be expressed in powers of $r$:
\[
\!\!\!\!\!\!\!\!\!\!\!
\vec{E}_\text{part}(\vec{r}, 0) &= \frac{q}{4\pi}\! \left\{
\frac{\hat{\vec{r}}}{r^2}-\frac{\vec{a}+(\vec{a}\cdot\rhat) \rhat}{2r}
+\bigl[\sfrac23\dot{\vec{a}}+\sfrac34(\vec{a}\cdot\rhat)\vec{a}+\sfrac38\left((\vec{a}\cdot\rhat)^2-a^2 \right)\hat{\vec{r}} 
\bigr]\right\} + O(r),
\\
\vec{B}_\text{part}(\vec{r},0) &= \frac{q}{8\pi}\hat{\vec{r}}\times\dot{\vec{a}}\, + O(r).
\]
In these expressions and below, time-dependent quantities such as $\vec{a}$ and $\dot{\vec{a}}$ 
are evaluated at $t = 0$.
These results are then used to compute the energy-momentum tensor \Eq{Tdef}.
Because we will integrate over the spherical shell at $r = r_0$, we compute
\[
\vec{T}\cdot\hat{\vec{r}}=&\frac{q^2}{16\pi^2 r^2}\left\{-\frac{\hat{\vec{r}}}{2r^2}+\frac{1}{2r}\left[\vec{a}+(\vec{a}\cdot\rhat)\rhat \right]+\sfrac12\left[ a^2-\sfrac14(\vec{a}\cdot\rhat)^2\right]\rhat-\sfrac54(\vec{a}\cdot\rhat)\vec{a}-\sfrac23\dot{\vec{a}} \right\}\nonumber\\
&-\frac{q}{4\pi r^2}\vec{E}_\text{ext}
+ O(r^{-1}).
\]
Performing the surface integrals in \Eq{depdt}, we find the simple result
\[
\eql{epresult}
\dot{\scr{E}}_\text{part} = O(r_0),
\qquad
\dot{\vec{p}}_\text{part} = \frac{q^2}{6\pi}\left(-\frac{\vec{a}}{r_0}+\dot{\vec{a}} \right)+q \vec{E}_\text{ext}
+ O(r_0).
\]
The fact that $\dot{\scr{E}}_\text{part} = 0$ can be understood from the fact
that the kinetic energy is quadratic in $\vec{v}$, and therefore the first order
change in the energy vanishes in the instantaneous rest frame.
Using $\dot{\vec{p}}_\text{part} = m_0 \vec{a}$
and taking $r_0 \to 0$, the momentum equation becomes
\[
\eql{ALrest}
\left( m_0 + \frac{q^2}{6\pi r_0} \right) \vec{a} = q \vec{E}_\text{ext}
+ \frac{q^2}{6\pi} \dot{\vec{a}}.
\]
The first term on the right-hand side is the Lorentz force for a particle 
in its rest frame, and the second is the Abraham-Lorentz `self force.'
We have put the term proportional to $\vec{a}$ on the left-hand side of the equation,
and we identify the combination
\[
\eql{massrenorm}
m = m_0 + \frac{q^2}{6\pi r_0}
\]
as the physical inertial mass of the particle.
The term $q^2 / 6\pi r_0$ represents the electromagnetic contribution to the mass of the 
particle, an effect first noticed by Thomson~\cite{Thomson1881}. 
It depends on $r_0$, which can be thought of as the size of
the particle in our simple model.

In a different regulator, we would find a different
result for the electromagnetic contribution to the mass, and therefore
a different relation between the `bare' mass $m_0$ and physical mass $m$.
However, all the model-dependence can be absorbed into the physical mass $m$,
so all regulators give the same `renormalized' equation of motion.
Note that the `counterterm' $q^2 / 6\pi r_0$ that is added to the mass $m_0$
to obtain a finite physical mass is divergent in the limit $r_0 \to 0$.
The radius $r_0$ is a `regulator' whose purpose is to make divergent quantities finite,
but the physical results are independent of $r_0$.
This `renormalization' procedure is conceptually similar to the one used in quantum 
field theory to eliminate short-distance divergences.
Note that if we take $r_0 \to 0$, we obtain $m_0 \to -\infty$.
The fact that the `bare' mass is negative will be important in the following.

Returning to \Eq{ALrest}, note that if we keep $r_0$ finite, we have additional terms 
in the equation of motion proportional to positive powers of $r_0$.
These are model-dependent, and represent corrections due to the structure of the
particle, as we will discuss later.
In the $r_0 \to 0$ limit we are considering here, these 
additional terms are not present.

We have derived the equation of motion \Eq{ALrest} in the rest frame of the particle,
but the equations of motion in a
general reference frame are determined by special relativity.
We can find the equation by writing a manifestly relativistically covariant equation
that reduces to \Eq{ALrest} in the rest frame of the particle.
This equation is unique and is given by
\[
\eql{AL}
m \frac{d u^\mu}{d\tau}
= q F^{\mu\nu}_\text{ext} u_\nu
+ \frac{q^2}{6\pi} \left[ \frac{d^2 u^\mu}{d\tau^2}
+ u^\mu \left( \frac{du^\nu}{d\tau} \frac{d u_\nu}{d\tau} \right) \right],
\]
where $\tau$ is the proper time, 
$F^{\mu\nu}_\text{ext}$ is the electromagnetic field strength tensor of the
external fields,
and $u^\mu = d X^\mu / d\tau$ is the 4-velocity of the particle.%
\footnote{A sketch of the derivation:
The relativistic equation can be written in terms of the 4-velocity $u^\mu(\tau)$.
In the instantaneous rest frame, we have
\[
u^\mu &= (1, \vec{0}),
\qquad
\frac{d u^\mu}{d\tau} = ( 0, \vec{a}),
\qquad
\frac{d^2 u^\mu}{d\tau^2} = (a^2,\,
\dot{\vec{a}}).
\]
Higher $\tau$ derivatives of $u^\mu$ bring in higher $t$ derivatives of $\vec{v}$
in the rest frame, and therefore cannot appear in the equations of motion.
It is straightforward to check that \Eq{AL} is the unique equation made from the 4-vectors
above that reduces to \Eq{ALrest} in the rest frame.}
In a general reference frame, the $\mu = 0$ component of this equation gives
a nontrivial energy conservation equation, and the spatial components of the
equation include the $\vec{v} \times \vec{B}$ term
in the Lorentz force, as well as relativistic corrections to the Abraham-Lorentz force.
\Eq{AL} was derived by Dirac \cite{Dirac:1938nz} using energy and momentum conservation
in a manifestly relativistic formalism.

\section{Energy and Momentum of Charged Particles\label{s.em}}
We now discuss the energy and momentum of a charged particle.
We will show that the model of a finite-size charged particle
used in the previous section suffers from a serious defect: 
the total energy and momentum of a single particle are not related
by the relativistic relation $\scr{E} = \sqrt{\vec{p}^2 + m^2}$.
However, this problem can be solved by including additional contributions
to the energy-momentum tensor inside the particle $(r < r_0)$, and that these
do not change the equation of motion derived in the previous section.

To see the problem,
let us compute the total energy and momentum of a charged particle
moving with constant non-relativistic  speed $\vec{v} = v \zhat$.
In the instantaneous rest frame, the contribution from the electromagnetic
fields is given by
\[
\scr{E}_\text{field} &= \int\limits_{r > r_0} d^3 r\, \sfrac 12 (E^2 + B^2)
= 4\pi \int_{r_0}^\infty r^2 \ggap dr \ggap \sfrac 12 \left( \frac{q}{4\pi} \right)^2
\frac{1}{r^4} + O(v^2)
= \frac{q^2}{8\pi r_0},
\\
p^z_\text{field} &= \int\limits_{r > r_0} d^3 r\ggap (\vec{E} \times \vec{B})^z
= 2\pi \int_{-1}^1 d\!\gap\cos\th
\int_{r_0}^\infty r^2 dr\ggap
\left( \frac{q}{4\pi} \right)^2 \frac{v(1 - \cos^2\th)}{r^4}
= \frac{q^2}{6\pi r_0} v.
\]
Note that the electromagnetic contribution to the energy differs from the contribution
to the momentum by a factor of $4/3$.
If we add the energy and momentum of the particle given by \Eq{Eppart}, the total
energy and momentum of the particle does not satisfy the relativistic relation.

This problem can be traced to another problem of this model, namely that the energy-momentum
tensor of the theory is not locally conserved, $\d_\mu T^{\mu\nu} \ne 0$.
The problem can be seen already for a charged particle at rest.
In this case we have (for $r > 0$)
\[
T^{00} &= \frac 12 \left( \frac{q}{4\pi} \right)^2 \frac{1}{r^4} \th(r - r_0),
\\
T^{0i} &= 0,
\\
T^{ij} &= -\left( \frac{q}{4\pi} \right)^2 \frac{\hat{r}^i \hat{r}^j - \frac 12 \de^{ij}}
{r^4} \th(r - r_0).
\]
The step functions $\th(r - r_0)$ encode the fact that the fields are nonzero
only for $r > r_0$.
We now check the local conservation of energy, $\d_\mu T^{\mu\nu} = 0$.
We find $\d_0 T^{00} + \d_i T^{i0} = 0$, but
\[
\d_0 T^{0i} + \d_j T^{ij} &= -\frac 12 \left( \frac{q}{4\pi} \right)^2 \frac{\hat{r}^i}{r_0^4}
\de(r - r_0).
\]
That is, local momentum conservation fails at $r = r_0$ due to the derivatives
acting on the step functions.
To obtain a result for the total energy that is compatible with relativity,
we must modify the energy-momentum tensor so that it is conserved, while
retaining Lorentz invariance.
A simple way to do this is to add a contribution that is non-vanishing
for $r < r_0$ in the rest frame:
\[
\eql{DeltaTmunu}
\De T^{\mu\nu} = \frac 12 \left( \frac{q}{4\pi} \right)^2 \frac{\eta^{\mu\nu}}{r_0^4}
\th(r - r_0),
\]
where $\eta^{\mu\nu}$ is the Minkowski metric.
The fact that this contribution is proportional the metric ensures that it
transforms as a tensor under Lorentz transformations, 
and therefore preserves the Lorentz invariance of the regulated theory.
With this addition, local conservation of energy and momentum is restored,
and we have (dropping $O(v^2)$ corrections)
\[
\scr{E}_\text{total} &= m_0 + \frac{q^2}{8\pi r_0} 
+ \int\limits_{r < r_0} d^3 r\ggap
\frac 12 \left( \frac{q}{4\pi} \right)^2 \frac{1}{r_0^4}
= m_0 + \frac{q^2}{6\pi r_0} = m,
\\
\vec{p}_\text{total} &= m_0 \vec{v} + \frac{q^2}{6\pi r_0} \vec{v}
= m \vec{v}.
\]
$\De T^{\mu\nu}$ does not contribute to $\vec{p}_\text{total}$ because
$\De T^{0i} \equiv 0$.
This restores the usual relation between energy and momentum.
In a general reference frame, the total energy and momentum is given 
by the 4-vector 
\[
p_\text{total}^\mu = m u^\mu,
\]
where $m$ is the same renormalized mass that
appears in the equation of motion \Eq{AL}.

The addition of the term \Eq{DeltaTmunu}
to the energy-momentum tensor does not affect the derivation of the
electromagnetic force in the previous section.
The reason is simply that if we take the sphere over which we integrate the
flux to be infinitesmally larger than $r_0$, 
the computation is unaffected by addition of $\De T^{\mu\nu}$.

Our choice for $\De T^{\mu\nu}$ is not unique.
We could have replaced our regulator by a physical model for an 
extended charge distribution, for example a spherical insulator of radius 
$r_0$ with charges on the surface.
Any such model would require non-electromagnetic forces to keep the charges
from flying apart due to the electromagnetic repulsion, and these forces
would give an additional contribution to the energy-momentum tensor
for $r < r_0$.
Lorentz invariant models that satisfy local energy-momentum conservation will
obviously produce energy and momentum that is compatible with Lorentz invariance.
(The importance of these additional contributions to the energy and momentum of
charged particles was first pointed out by Poincar\'e \cite{Poincare1905}.)
Our choice of $\De T^{\mu\nu}$ is made for simplicity, since
the details of how we model the region $r < r_0$ are not
important in the limit $r_0 \to 0$.
The important point is that the energy-momentum tensor must be compatible
with relativity and local conservation of energy and momentum in order to 
obtain the correct relation between energy and momentum of the particle.

\section{Classical Electrodynamics of Point Particles...Isn't\label{s.paraAL}}
In this section we work out the consequences of the AL equation in the limit
of point particles ($r_0 \to 0$).
Before we do this, we note that elementary considerations show that
classical electrodynamics in this limit is subject to catastrophic
instabilities.
Consider a configuration consisting of a pointlike electron and positron.
When the particles are far apart, the energy of this configuration is $2m$,
the rest mass energy.
When the particles are a distance $d$ apart, the energy is
\[
\scr{E} = 2m - \frac{e^2}{4\pi d},
\]
provided that the particles are at rest.
This can be made arbitrarily negative by making $d$ arbitrarily small,
so the total energy is unbounded from below.
This means that energy conservation allows an infinite amount of energy
to be radiated as the particles accelerate toward each other.
This radiation energy can be collected, allowing the creation of a perpetual
motion machine.
In our world, we are saved from this catastrophe by quantum mechanics:
the uncertainty principle does not allow particles to come arbitrarily close 
to each other.
The minimum distance is given by
\[
d_\text{min} \sim \frac{4\pi \hbar^2}{q^2 m},
\]
the analog of the Bohr radius for positronium.

We now show that in the point particle limit, radiation reaction implies
another catastrophic instability:
the generic behavior of charged particles
is a rapid `self-acceleration' to nearly the speed of light.
(We will show below that in our world this instability is eliminated by
\emph{relativistic} quantum mechanical effects.)

The self-accelerated behavior can be seen in solutions of the AL
equations in the absence of external forces.
We begin by reviewing these well-known `self-accelerated' solutions.
Because the solution predicts that the particle rapidly
approaches the speed of light, we give the relativistic form of this
solution.
We consider a point particle in the absence of external fields.
We parameterize the particle's 4-velocity as 
$u^\mu=(\dot{T}(\tau),\dot{\vec{X}}(\tau))$.
(We now use dots to denote derivatives with respect to proper time.)
We obtain
\[
\eql{runawayeq}
\ddot{T} &= \tc \left[\dddot{T} +\dot{T}\left(\ddot{T}^2-\ddot{X}^2 \right)\right],
\qquad
\ddot{\vec{X}} = \tc \left[\dddot{\vec{X}} +\dot{\vec{X}}\left(\ddot{T}^2-\ddot{X}^2 \right)\right],
\]
where
\[
\eql{tauc}
\tc = \frac{q^2}{6\pi m}.
\]
To obtain an explicit solution, consider a particle trajectory along the $x$ axis.
The 4-velocity satisfies the constraint $u^2=1=\dot{T}^2-\dot{X}^2$,
which implies $\dot{T}\ddot{T} = \dot{X} \ddot{X}$, and therefore
\[
\ddot{X}^2=\frac{\dot{T}^2\ddot{T}^2}{\dot{T}^2-1}.
\]
We can then write the first equation in \Eq{runawayeq} as
\[
\ddot{T}= \tc \left(\dddot{T} -\frac{\dot{T}\ddot{T}^2}{\dot{T}^2-1}\right).
\]
Straightforward integration leads to the solution
\[
\dot{T}=\cosh\left(c_1 e^{\tau/\tc} +c_2 \right),
\qquad
\dot{X}=\sinh\left(c_1e^{\tau/\tc}+c_2 \right),
\]
which means the 3-velocity is
\[
v(\tau)=\frac{\dot{X}}{\dot{T}}=\tanh\left(c_1 e^{\tau/\tc}+c_2 \right).
\]
If we assume that the particle was at rest at $\tau \to -\infty$, we have
$c_2 = 0$.
The interpretation of such a solution is that in the far past the particle was
infinitesmally perturbed, and then began to self-accelerate.
Then $c_1 = \tanh^{-1}(v_0) = \eta_0$, where $v_0$ (respectively $\eta_0)$ is the 
3-velocity (respectively rapidity) of the particle at $\tau = 0$.
At late times $\tau \gg \tc$ we have 
\[
\ga = (1- v^2)^{-1/2} = \sfrac 12 e^{\frac 12 \eta_0 e^{\tau/\tc}}.
\]
That is, the boost factor $\ga$ of the particle
is increasing as the exponential of an exponential!%
\footnote{In terms of coordinate time, we have $d\tau = 2 e^{-t/\tc}$
for $\tau \gg \tc$, and therefore
\[
\frac{d\ga}{dt} = \frac{\eta_0}{2\tc} 
e^{\frac 12 \eta_0 e^{\tau/\tc}},
\]
so the boost factor is increasing super-exponentially fast in terms of
coordinate time as well.}

We now show that the runaway solutions are unavoidable if we require that the theory 
is causal, meaning that the past behavior of the universe predicts its future.
Let us consider a particle that is at rest in the past, and is then subject to
an external force that acts for a finite time interval $t_i \le t \le t_f$.
For simplicity, we consider non-relativistic motion in one dimension, 
where the equation we want to solve is
\[
\eql{ALFext1D}
\frac{d^2}{dt^2} X(t) - \tc \frac{d^3}{dt^3} X(t)
= \frac{F(t)}{m},
\]
with initial conditions $X(t) \equiv 0$ for $t < t_i$.
This is a linear equation in $X(t)$, so we can
solve it for an arbitrary external force $F(t)$ using a Green's function:
\[
X(t) = \int_{-\infty}^\infty dt' \ggap G(t - t') \frac{F(t')}{m},
\]
where
\[
G(t) = \begin{cases}
0 & t < 0, \\
-\tc \left[ e^{t/\tc} - \frac{t}{\tc} - 1 \right] & t > 0.
\end{cases}
\]
Note that we have incorporated causality because the Green's function $G(t - t')$
is nonzero only for $t > t'$.
The particle will self-acclerate for $t > t_f$ unless $\ddot{X}(t_f) = 0$,
but we have
\[
\ddot{X}(t_f) = -\frac{1}{m \tc} \int_{t_i}^{t_f} dt\ggap e^{-(t - t_f)/\tc}
F(t).
\]
This is nonzero for a generic external force, so runaway
behavior is physically inevitable.
Some authors have advocated prescriptions that can eliminate
the runaway behavior (see for example \cite{Dirac:1938nz}),
but these violate causality and will not be discussed here.

The runaway solutions are exact solutions to the AL equation, which was
derived from conservation of energy and momentum, and so energy and momentum
is conserved in these solutions.
This appears paradoxical, since a self-accelerated charge radiates energy
and momentum.
In fact the power radiated is given by Li\'{e}nard's relativistic generalization
of the Larmor formula, and increases exponentially with proper time:
\[
\eql{selfrad}
P_\text{rad} = 
-\frac{q^2}{6\pi} \dot{u}^\mu \dot{u}_\mu
= \frac{m \eta_0^2}{\tc} e^{2\tau / \tau_c}.
\]
The resolution of this apparent paradox lies in the fact that the `bare'
mass of the particle is negative for $r_0 \to 0$.
To see this, we consider the energy and momentum transferred to the field
outside the shell surrounding the particle in its instantaneous rest frame.
For this, we have to include the term proportional to $\vec{a}$ in \Eq{epresult}.
That is, in the rest frame we have (for vanishing external fields)
\[
\eql{epresult}
\dot{\scr{E}}_\text{field} = -\dot{\scr{E}}_\text{in} = 0,
\qquad
\dot{\vec{p}}_\text{field} = -\dot{\vec{p}}_\text{in}
= \frac{q^2}{6\pi}\left(\frac{\vec{a}}{r_0}-\dot{\vec{a}} \right),
\]
where `field' refers to the energy/momentum in the fields for $r > r_0$,
and `in' refers to the energy/momentum for $r < r_0$.
Here a dot again denotes a derivative with respect to coordinate time.
If $r_0$ is smaller than any other scale in the problem, the term proportional
to $1/r_0$ in \Eq{epresult} dominates.
The covariant generalization of the leading result is then
\[
\frac{d p_\text{field}^\mu}{d\tau} = \frac{\tc}{r_0}
\frac{du^\mu}{d\tau}.
\]
In a general reference frame, we then have
\[
\dot{\scr{E}}_\text{field} &= \tc \ga^3 \frac{\vec{v} \cdot \vec{a}}{r_0},
\qquad
\dot{\vec{p}}_\text{field} = \tc \ga 
\frac{\vec{a} + \ga^2 (\vec{v} \cdot \vec{a}) \vec{v}}{r_0}.
\]
We see that for $\vec{v} \cdot \vec{a} > 0$, the energy in the field is increasing,
as we expect for a particle that is radiating.
We also have
\[
\dot{\vec{p}}_\text{field} \cdot \vec{v} = \frac{\ga^3 \tc}{r_0} \vec{v} \cdot \vec{a},
\]
so the field momentum along the direction of motion of the particle is also increasing,
again as we expect for a particle that is radiating.
So why is the particle speeding up rather than slowing down?
The reason is that the energy and momentum of the particle (everything inside the
shell) in the rest frame is given by
\[
\eql{min}
\scr{E}_\text{in} = m - \frac{q^2}{8\pi r_0} \equiv m_\text{in},
\qquad
\vec{p}_\text{in} = 0,
\]
and therefore in a general frame by
\[
\scr{E}_\text{in} = \ga m_\text{in},
\qquad
\vec{p}_\text{in} = \ga m_\text{in} \vec{v}.
\]
Note that we have included the contribution to the energy-momentum tensor
that ensures local conservation of energy and momentum at the boundary,
\Eq{DeltaTmunu}.
The important point is that $m_\text{in} < 0$ in the limit $r_0 \to 0$,
meaning that everything inside the radius $r = r_0$ has negative mass.
Negative mass means that the kinetic energy of the particle gets smaller (more negative)
as the speed increases, and the momentum is in the opposite direction to the
velocity.
In other words, the 
self-accelerating particle conserves energy and momentum because its
contribution to the energy and momentum is opposite from that of an ordinary
positive mass particle.
The importance of the negative bare mass in understanding the conservation
of energy and momentum in the self-accelerated solutions was (to our knowledge)
first emphasized by Coleman~\cite{Coleman:1961zz},
although it is implicit in the earlier work of Dirac~\cite{Dirac:1938nz}.
General aspects of the physics of negative mass particles are discussed in \cite{Price}.

The theory with $r_0 \to 0$ clearly does not describe what we observe
in nature, but the way out is clear:
we must abandon the assumption that $r_0$ is the smallest scale in the
problem.
If we assume $r_0 \gg \rc$, then the energy inside a sphere of
radius $r_0$ is positive (see \Eq{min}), and there is no negative
energy inside the particle to power the self-accelerated solutions.
Indeed, detailed analysis of physical models of finite-size charges 
confirms the absence of runaway behavior if
the physical size of the charge is large compared to $\rc$ \cite{Sommerfeld1905}.

In fact, physical elementary particles such as the electron have 
an effective size due to relativistic quantum effects, such that
the classical description of elementary particles such as the electron breaks down at
a distance scale much larger than $\rc$.
If we attempt to localize an electron on a sufficiently small spatial region, the 
uncertainty principle implies that this
region will contain enough momentum (and therefore energy) to create 
particle-antiparticle pairs, invalidating the classical description.
This occurs for length scales smaller than the quantum length scale
(temporarily restoring factors of $c$)
\[
r_\text{Q} = \frac{\hbar}{m c}.
\]
(This scale is related to the Compton wavelength, $\la_\text{Compton} = 2\pi r_\text{Q}$.)
In fact, we have
\[
\frac{\rc}{r_\text{Q}} = \frac{e^2}{4\pi\hbar c} = \al,
\]
where $e$ is the charge of the electron
and $\al \simeq \frac{1}{137}$ is the fine-structure constant.
Since the classical description breaks down at a radius much larger than $\rc$,
we must take the size of the electron to be
$r_0 > r_\text{Q} \gg \rc$ to justify the use of classical physics.

\section{The Effective Abraham-Lorentz equation}
We have argued above that if we want to avoid the instabilities that occur for point
charged particles, we must assume that the size of the classical particle is
$r_0 \gg \rc$.
Away from the $r_0 \to 0$ limit, it may seem that
we must give up the simplicity of the point particle description.
In this section we will show that one can systematically incorporate the requirement
$r_0 \gg \rc$ into the point particle approximation.
We are interested in situations where the physical size of the particle is much smaller 
than the other length and time scales in the problem.
In this case, we can treat the particle as a structureless classical 
point particle to first approximation.
The effects of the structure of the particle can then be included systematically
as a series of corrections to the point particle limit, as in the multipole
expansion for the electromagnetic field of a small charge distribution.
This is an example of an `effective theory' that describes physics at
long distances and times in a systematic expansion in short-distance structure.
Such theories are an important idea in modern condensed matter 
and elementary particle physics.

We begin by noting that for $r_0 \gg \rc$, the bare mass $m_0$,
the renormalized mass $m$, and the `inside' mass $m_\text{in}$ are all
approximately equal since (see \Eqs{massrenorm} and \eq{min})
\[
m, m_\text{in} = m_0 + O(\rc/r_0).
\]
In particular, all of these masses are positive, and any of them may be taken
to the `the' mass of the particle to good approximation.
The discussion of the previous section then tells us that the runaway instability should
be absent (since in particular $m_\text{in} > 0$).

Even though $r_0$ is no longer the smallest scale in the problem,
we expect that we can use the point particle approximation as
long as the characteristic length $L$ and time $T$ scales of the motion are much 
larger than the size of the particle.
The meaning of $L$ and $T$ is that in the instantaneous rest frame
$|\vec{a}| \sim L/T^2$,  $|\dot{\vec{a}}| \sim L/T^3$, {\it etc\/}.%
\footnote{More precisely,
dimensional analysis allows us to write the trajectory of the particle as
$\vec{X}(t) = L \gap \vec{f}(t/T)$, where $\vec{f}$ is a dimensionless function
of a dimensionless argument.
We assume that $L$ and $T$ are such that $|\vec{f}(s)| \lsim 1$ as long
as $|s| \lsim 1$.}
In other words, we are interested in the approximation
\[
\eql{hierarchy}
\rc \ll r_0 \ll L, T.
\]
In this approximation, the expansion in powers of $r_0$ that leads to the
equation of motion is a series in parametrically smaller terms.
Including higher order terms in $r_0$ neglected in the derivation in the previous section,
the equation of motion for a charged partice subject to an external force
$\vec{F}_\text{ext}$ is given by (in the instantaneous rest frame)
\[
\eql{aexpand}
\vec{a} = \underbrace{\frac{\vec{F}_\text{ext}}{m}}%
_{\displaystyle{}\sim \, \frac{L}{T^2}}
\, + \, \underbrace{\frac{q^2}{6\pi m} \dot{\vec{a}}}%
_{\displaystyle{}\sim \, \frac{\rc L}{T^3}}
\, + \underbrace{C_1 \frac{q^2}{6\pi m} r_0 \ddot{\vec{a}}}%
_{\displaystyle{}\sim \, \frac{\rc r_0 L}{T^4}}
\, + \underbrace{C_2 \frac{q^2}{6\pi m} r_0 a^2\vec{a}}%
_{\displaystyle{}\sim \, \frac{\rc r_0 L^3}{T^6}}
+ \underbrace{C_3 \frac{q^2}{6\pi m} r_0^2 \gap \dddot{\vec{a}}}%
_{\displaystyle{}\sim \, \frac{\rc r_0^2 L}{T^6}}
+ \cdots\,
\]
where the $C_i$ are order-one, model-dependent, dimensionless coefficients. 
The pattern is that the radiation reaction terms are all suppressed by one
power of $\rc$, and the model-dependent corrections are suppressed by additional
powers of $r_0$.
Because of the hierarchy of length scales \Eq{hierarchy} this can be thought of as an 
expansion in powers of the small parameters $\rc$ and $r_0$.

Motivated by the expansion above, we can substitute the approximation
\[
\eql{adotapprox}
\dot{\vec{a}} = \frac{d}{dt} \left( \frac{\vec{F}_\text{ext}}{m} \right)
+ O\!\left( \frac{\rc L}{T^4} \right)
\]
into \Eq{aexpand}
to write the approximate equation of motion
\[
\eql{ALeff}
\vec{a} \simeq \frac{\vec{F}_\text{ext}}{m}
+ \tc  \frac{d}{dt} \left( \frac{\vec{F}_\text{ext}}{m} \right).
\]
In the literature, this is called the `reduced order' AL equation because it is a second
order differential equation (unlike the original AL equation, which is 
third order in time derivatives).
This approximation is not new \cite{Eliezer1948,Landau1948},
but what appears to be missing in the literature is a clear explanation
of why we should use the approximation \Eq{adotapprox} rather than
simply dropping the model-independent corrections (depending on $r_0$)
in \Eq{aexpand}, yielding the AL equation written in the previous section.
The reason is simply that the terms omitted in \Eq{adotapprox} give corrections
to the equation of motion that are suppressed compared to the corrections from the
model-dependent terms in \Eq{aexpand}.
This is because the corrections to \Eq{adotapprox} are suppressed by an additional
power of the small parameter $\rc$.
For example, 
the ratio of the correction from \Eq{adotapprox} to the correction to the leading
model-dependent correction is of order
\[
\tc \frac{\rc L}{T^4} \bigg/ 
\frac{\rc r_0 L}{T^4}
\sim \frac{\rc}{r_0} \ll 1.
\]
We see that the corrections from the omitted terms are less important
than the model-dependent corrections to \Eq{aexpand}.
We will see that \Eq{ALeff} is free of runaway solutions
or other pathologies.
In other words, for $r_0 \gg \rc$, the runaway solutions of the previous
section are invalidated by model-dependent corrections.
This makes good physical sense, since the model-dependent terms 
contain the information about the finite size of the particle, 
which we argued above allows the particle to have positive mass,
and therefore no instabilities.

We see that we can think of \Eq{ALeff} as an `effective theory' that gives
the leading approximation to radiation reaction in an expansion in powers of $r_0$.
We therefore call \Eq{ALeff} the `effective AL equation.'
If desired, we could systematically improve the approximation by including
terms with higher powers of $r_0$.
However, these corrections will depend on additional unknown coefficients
(the $C_i$ in \Eq{aexpand})
that parameterize the short-distance structure of the particle.
This kind of effective theory 
expansion in a powerful tool of modern theoretical physics.

Because $r_0$ is no longer the smallest scale in the problem, we must also reconsider
the expansion of the external fields in \Eq{extfieldexpand}.
At $t = 0$ in the instantaneous rest frame we can use the Taylor expansion
\[
\vec{E}_\text{ext}(\vec{r}) = \vec{E}_{\text{ext}}(0)
+ r^i \partial_i \vec{E}_{\text{ext}}(0,0)
+ \sfrac 12 r^i r^j \partial_i \partial_j \vec{E}_{\text{ext}}(0)
+ O(r^3).
\]
The cross terms between $\vec{E}_\text{ext}$ and $\vec{E}_\text{part}$ give rise to
corrections to the external force that begin at $O(r_0^2)$:
\[
\eql{gradientforces}
\dot{\vec{p}}_\text{part}
&= q \vec{E}_\text{ext}(0) 
+ \frac{q}{2} r_0^2 \nabla^2 \vec{E}_\text{ext}(0)
\nonumber\\
&\qquad{}
- \frac{q}{6}  r_0^2 \Bigl[ \vec{a} (\bm{\nabla} \cdot \vec{E}_\text{ext})
- \vec{a} \times (\bm{\nabla} \times \vec{E}_\text{ext})
+ (\vec{a} \cdot \bm{\nabla}) \vec{E}_\text{ext} 
\Bigr] + \cdots
\]
We assume that the external fields vary on length scales $L_\text{ext}$.
Requiring the $O(r_0^2)$ corrections to be small compared to the AL term requires
\[
\eql{externalfield}
L_\text{ext} \gg \frac{r_0^2 L}{\rc T},\ggap
\left( \frac{r_0^2 T}{\rc} \right)^{1/2}.
\]
If either of these conditions is violated, the gradient forces in \Eq{gradientforces}
are more important for the instantaneous motion of the charged particle than
the AL force term.
If this is the case, the structure of the charged particle cannot be neglected, 
and we cannot use the point particle approximation.

\subsection{General Consequences of the Effective Abraham-Lorentz Equation\label{s.paraEAL}}
The final result we have derived above is very simple and intuitive.
The effective AL equation \Eq{ALeff} is a second-order differential equation,
where radiation reaction effects are parameterized by a term suppressed by 
a small coefficient $\tc$.
The fact that the equation is second order means that the trajectory
is determined by the usual initial conditions, for example the initial
position and velocity of the particle.
It is also easy to see that the self-accelerated solutions are absent, since
the acceleration of the particle vanishes if the external force vanishes.

The effective AL equation was derived assuming conservation of energy and momentum.
In situations where there is a well-defined energy carried away by electromagnetic
radiation, the effective AL equation reproduces what we expect from simple
energy conservation conditions.
Neglecting radiation reaction, the work done on the particle by the external force
is given by 
(assuming non-relativistic motion for simplicity)
\[
W_\text{ext} = \int_{t_i}^{t_f} \!\! dt\, \dot{\vec{X}} \cdot
\vec{F}_\text{ext}.
\]
The work done on the particle by the radiation reaction term is given by
\[
W_\text{rad} &= \int_{t_i}^{t_f} \!\! dt\, \dot{\vec{X}} \cdot
\tc \frac{d}{dt} \left( \frac{\vec{F}_\text{ext}}{m} \right)
\nn
&= \frac{\tc}{m} \left[ \dot{\vec{X}} \cdot \vec{F}_\text{ext} 
\right]_{t \, = \, t_i}^{t \, = \, t_f}
- \tc \int_{t_i}^{t_f} \!\! dt\, \ddot{\vec{X}} \cdot 
\frac{\vec{F}_\text{ext}}{m},
\]
where we used integration by parts in the second line.
The first term vanishes in many cases of interest.
For example, it vansishes if the external force vanishes at early and late times,
and it vanishes if the applied force is periodic (provided that $t_f - t_i$ is
chosen to be a single period).
In these cases, we have
\[
W_\text{rad} =  -\tc \int_{t_i}^{t_f} \!\! dt\, |\ddot{\vec{X}}|^2 
+ O(\rc^2).
\]
The integral on the \rhs\ is the total energy radiated by the charge,
as computed by the Larmor formula.
That is, up to negligibly small corrections of order $\rc^2$, 
energy conservation reduces to the intuitive statement that
the change of the kinetic energy of the particle in the time interval $t_i < t < t_f$ 
is given by the work done by the external fields minus the energy radiated.
A version of this argument first appeared in \cite{Lorentz1906}.
It is not hard to check that the argument above can be extended to the relativistic
case, and we leave this as an exercise for the reader.

\subsection{Constant Acceleration}
Although the effective AL equation has many nice features, it shares a counter-intuitive
prediction with the original AL equation, namely that a particle with constant
acceleration has no radiation reaction, even though the Larmor formula
predicts that such a particle radiates energy to infinity.
In this subsection, we make some brief comments on this famous question.

First, note that the simple conservation of energy argument made in the previous
subsection does not apply to the case of strictly constant acceleration
because such a particle is never in inertial motion.
We therefore consider a charged particle that experiences a constant 
acceleration in a finite time interval, with inertial motion before and
after.
As a concrete example, we consider a charged particle moving through an infinite 
plane capacitor with width $d$ and internal electric field 
$\vec{E}_\text{cap} = E_\text{cap} \zhat$.
In this setup, the general arguments of the previous subsection apply, and we
are guaranteed that energy radiated will match the loss of kinetic energy of the
particle (compared to the situation where we neglect radiation reaction).
It is instructive to see how this comes about in a concrete
example.
Note that the particle radiates at a constant rate while inside the capacitor, 
so the energy radiated grows with $d$, while the effective AL equation predicts 
that the particle experiences a radiation reaction force only as it enters and 
leaves the capacitor.
It is not immediately obvious that this can work, and in fact we will find
some unusual aspects of radiation reaction in this example.

We consider for simplicity non-relativistic motion in the $z$ direction.
The effective AL equation in this case is
\[
\ddot{Z} 
&= a_\text{cap} \th(Z) \th(d-Z)
+ \tc a_\text{cap} \bigl[ \de(Z) - \de(Z - d) \bigr] \dot{Z},
\eql{effcapeq}
\]
where $a_\text{cap} = q E_\text{cap}/m$ is the acceleration of the particle inside
the capacitor.
For $t < 0$ we have
\[
Z(t) = v_i t.
\]
At $t = 0$ the particle reaches the capacitor.
The delta function term in \Eq{effcapeq} causes the velocity of the
particle to jump at $t = 0$, so while the particle is inside the capactor
the solution is
\[
Z(t) = v_\text{cap} t + \sfrac 12 a_\text{cap} t^2,
\]
where
\[
v_\text{cap} = v_i + a_\text{cap} \tc.
\]
Note that the radiation reaction causes the particle to speed up if $a_\text{cap} > 0$;
we will comment on this below.
The particle then stays in the capacitor for a time $T$ given by
\[
T = \frac{1}{a_\text{cap}} \left[ -v_\text{cap} + \sqrt{v_\text{cap}^2 + 2 a_\text{cap} d} \, \right].
\]
At $t = T$, the velocity of the particle again jumps, and we have for $t > T$
\[
Z(t) = v_f t,
\]
where
\[
v_f = v_\text{cap} + a_\text{cap} T - a_\text{cap} \tc = v_i + a_\text{cap} T.
\]
Expanding in the small parameter $\tc$ we obtain
\[
\eql{Tcorr}
T = T_0 - \tc \left[ 1 - \sqrt{1 + \frac{2 a_\text{cap} d}{v_i^2}}\, \right]
+ O(\tau_\text{c}^2),
\]
where
\[
T_0 = \frac{1}{a_\text{cap}} \left[ -v_i + \sqrt{v_i^2 + 2 a_\text{cap} d} \, \right]
\]
is the time the particle would spend in the capacitor if we neglect
radiation reaction.
Note that the argument of the square root in \Eq{Tcorr} is positive provided that
the particle makes it out of the capacitor.
Note that the time difference due to radiation reaction
is extremely small, since it is proportional to $\tc$.
The change in the kinetic energy in the particle can then be found to be
\[
\sfrac 12 m v_f^2 - \sfrac 12 m v_i^2
= q E d + m \tc a_\text{cap}^2 T.
\]
The first term on the \rhs
is the work done by the capacitor on the particle, and the second
is the energy lost to radiation as predicted by the Larmor formula.
This agrees with the general argument given in the previous subsection.

An apparently counterintuitive feature of this solution is that the radiation
reaction causes the particle to speed up as it enters the capactor for
$a_\text{cap} > 0$.%
\footnote{It would be interesting to observe this effect experimentally, but it is
very small because it is of order $\tc$.}
We might wonder where the energy for this boost comes from.
The only possible answer is that it comes from \emph{reducing} the energy
stored in the electric field of the capacitor.
The energy density due to the electromagnetic field is given by
\[
u = \sfrac 12 (\vec{E}_\text{cap} + \vec{E}_\text{part})^2 + \sfrac 12 \vec{B}_\text{part}^2,
\]
where $\vec{E}_\text{part}$, $\vec{B}_\text{part}$ denote the fields of the particle.
This contains a cross term $\vec{E}_\text{cap} \cdot \vec{E}_\text{part}$, which
can be negative.
Checking that the energy balance works in this case at intermediate times is neither
simple nor instructive, so we omit it.
It is guaranteed to work in any case, since the effective ALD equation was
derived by assuming that the transfer of energy between the particle
and the field conserves the total energy.

\section{Conclusions\label{s.con}}
We have shown that radiation reaction in classical electromagnetism
can be understood by a systematic expansion in powers of $r_0$, the size of
a charged particle.
In the point particle limit $r_0 \to 0$ the theory suffers from catastrophic
instabilities:
the energy is unbounded from below, and
charged particles rapidly `self-accelerate' to nearly the speed of light.

These instabilities are absent if the size of the particle is larger than the
particle's classical radius $\rc$, the radius at which the electromagnetic contribution
to the particle's mass is equal to its total mass.
In our world, the classical radius of all particles is smaller than the scale
at which quantum effects become important, so the classical theory is invalid
in the regime where the instabilities appear.

If $r_0 \gg \rc$, but $r_0$ is much smaller than other scales in the problem,
we can find the effects of radiation reaction in a systematic expansion in powers
of $r_0$.
The leading term in this expansion gives the `effective Abraham-Lorentz equation,'
a second order differential equation that does not have instability problems
and which incorporates energy conservation in a simple way.
We illustrated the use of this equation with the case of constant acceleration
in a finite time interval.
This expansion can be improved systematically by including additional terms;
it is an example of an `effective theory,' an important tool in modern
theoretical physics.

We do not wish to claim any major new results in our treatment.
In fact, essentially everything in this paper appears somewhere in the early 
literature on the subject.
However, we have given a unified treatment from a modern point of view,
and we hope that this will help clarify some aspects of 
this notoriously confusing and classic problem
for both students and researchers in physics.

\section*{Acknowledgements}
We thank Ira Rothstein,
James Scargill,
and John Terning
for helpful discussions. 
M.L. and C.B.V. are supported by Department of Energy grant No. DE-SC-0009999.

\newpage\frenchspacing
\bibliographystyle{utphys}
\bibliography{Radbib}

\end{document}